\documentclass[a4paper]{jpconf}
\usepackage{graphicx}
\usepackage{mhchem}
\graphicspath{images/}
\usepackage{siunitx}
\usepackage{epstopdf}
\DeclareSIUnit\sq{\ensuremath{\opensquare}}

\begin{document}
\title{Performance of a fast timing micro-pattern gaseous detector for future collider experiments}

\author{Antonello Pellecchia, Piet Verwilligen}

\address{Università di Bari \& INFN sezione di Bari, Via E. Orabona 4, 70125 Bari, Italy}

\ead{antonello.pellecchia@cern.ch}

\begin{abstract}
The fast timing MPGD is a micro-pattern gaseous detector conceived for achieving sub-nanosecond time resolution while maintaining the ability to instrument large areas in high-rate environments; applications of such technology are perspected in high-energy physics experiments at future colliders and medical diagnostics with time-of-flight methods. This work shows the systematic studies carried on an FTM prototype on the performance of GEM foils coated with resistive DLC films, whose development is essential for the FTM operation. The resistive foil performance has been tested with several gas mixtures and compared with the results obtained on conductive foils. The results show that the performance of the FTM is presently limited by the technology of manufacturing of DLC-coated GEM foils, with high gains reachable exclusively in isobuthane-based mixtures.
\end{abstract}

\section{Introduction}

The achievement of precise timing has emerged as a technological requirement for the design of high-energy physics experiments at future colliders as a necessary tool for pile-up mitigation and particle identification with time-of-flight methods. The class of micro-pattern gaseous detectors (MPGDs) includes consolidated technologies for the instrumentation of large areas in high-energy physics experiments in environments of rates higher than \SI{100}{\kilo\Hz\per\centi\m^2}, with excellent space resolutions under \SI{100}{\micro\m} and time resolutions of the order of \SI{5}{\nano\s}; the fast timing MPGD (FTM) is a gaseous detector designed for the achievement of a sub-nanosecond time resolution for precise timing applications over large areas in future collider experiments.

\subsection{Working principle of the FTM}


The geometry of a traditional MPGD can be divided in a drift region and an amplification region. 
The main contribution to a traditional MPGD time resolution is given by the fluctuations in the position of the primary ionization cluster created closest to the amplification region by a charged particle passing through the gas; it can be quantified as $\sigma_t^{(1)} = (\lambda v_d)^{-1}$, where $v_d$ is the average electron drift velocity in the drift gap and $\lambda$ is the average number of primary clusters created by the particle per unit length.


In the FTM, the detector active volume is divided in a stack of layers each with its own drift gap and amplification region \cite{rui_ftm}; the main contribution to the detector time resolution is now the fluctuation in the position of the primary cluster that is created closest to \emph{any} amplification region. Its time resolution can then be parameterized as $\sigma_{t}^{(N)} = (N\lambda v_d)^{-1} = \sigma_t^{(1)}/N$.

An essential detail in the FTM design is that the detector should be electrically transparent throughout all the layers to allow signal induction on an external readout electrode: all the other electrodes of the FTM should then be resistive. The progress in the techniques for the production of resistive foils for the instrumentation of MPGDs is then an essential requisite for the development of the FTM. The measurements performed on the FTM prototype presented in the following sections were aimed at determining the performance of the resistive GEM foils available with the present state-of-the-art production technology towards the construction of a detector suitable for the operation with an high number of layers.

\section{The small-size FTM prototype}

\begin{figure}
    \centering
    \begin{minipage}[b]{.5\textwidth}
        \includegraphics[width=.9\textwidth]{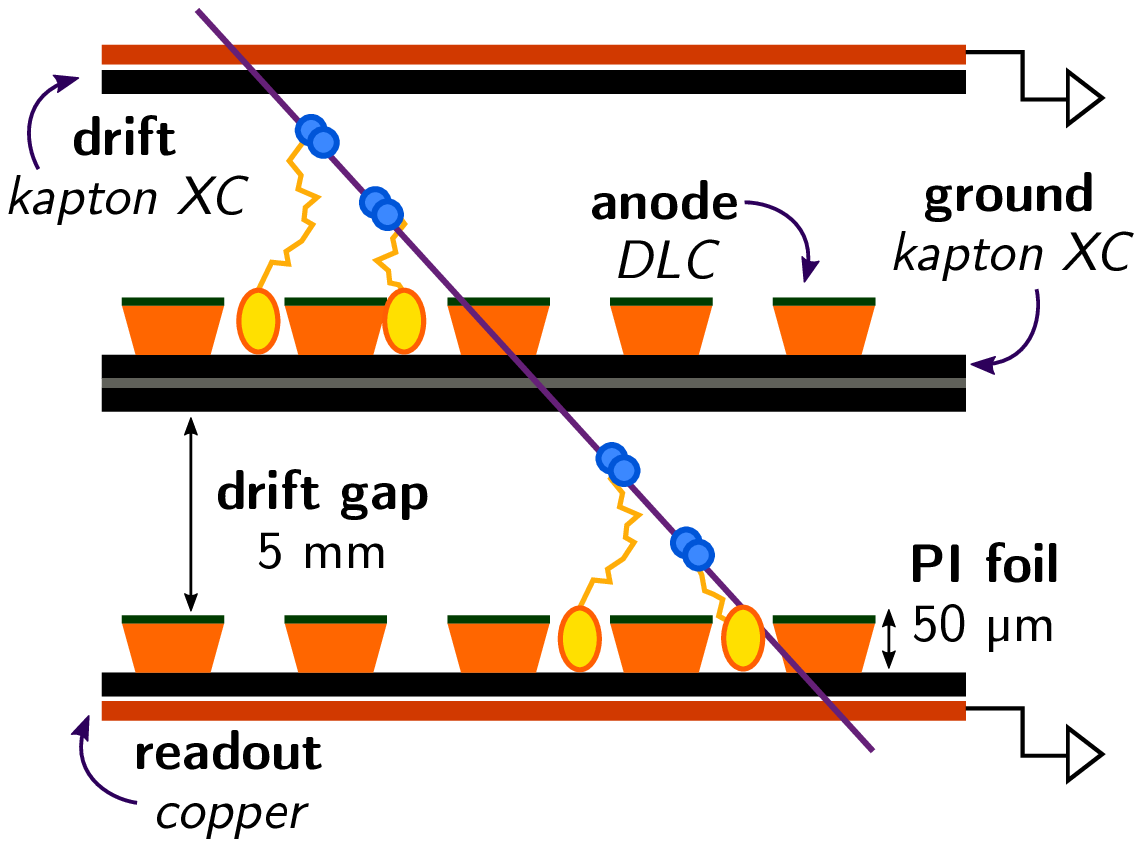}
    \end{minipage}
    \begin{minipage}[b]{.45\textwidth}
        \includegraphics[width=.9\textwidth]{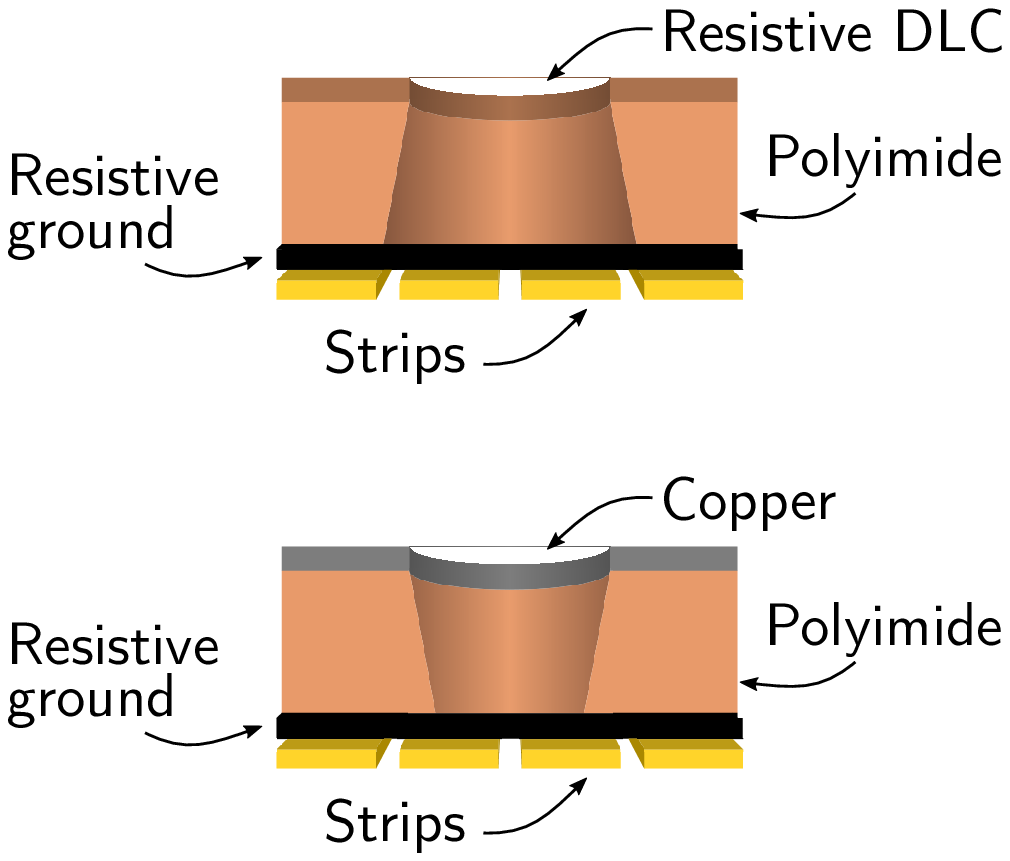}
    \end{minipage}
    \caption{On the left, structure of the small-size FTM prototype instrumented with two layers. On the right, comparison between the FTM amplification foils and the foils used in the µ-RWELL detector, which differ for the copper electrode as well as in the inverted hole shape.}
    \label{fig:ftm_2layers}
\end{figure}

The small-size FTM prototype has been conceived with a modular design to be operated with a variable number of layers. The structure of the prototype instrumented with two layers is shown in the left side of Fig.~\ref{fig:ftm_2layers}. Each layer is delimited by a \emph{drift} electrode on the top and a \emph{ground} electrode on the bottom, both made by a thin layer of resistive polyimide of surface resistivity of approximately \SI{5}{\mega\ohm\per\sq}. The amplification foil is a resistive well, made of a \SI{50}{\micro\m}-thick polyimide layer covered by a \SI{100}{\nano\m}-thick layer of diamon-like carbon (DLC, the \emph{anode} electrode). The DLC was deposited on the polyimide by magnetron sputtering and from the resulting laminate the GEM foil was obtained by chemical etching; the anode surface resistivity is of the order of \SI{100}{\mega\ohm\per\sq} before the etching. Two external readout electrodes of \SI{1}{\centi\m} radius enclose the hole structure, allowing signal pickup from any layer. The active surface of the entire detector is \SI{2}{\centi\m\squared}. The geometry of a single layer of the small-size FTM is similar to a µ-RWELL detector, with the exception of the resistive anode and the inverted hole shape (right side of Fig.~\ref{fig:ftm_2layers}).

\section{Tests on the FTM prototype with laser setup}

The prototype has been tested in an optical setup using a UV laser as a source designed for gain and efficiency measurements. Beside allowing to test one layer of the detector at a time, laser beams can release a fixed energy in detectors of small gas volumes, while the amount of charge created by ionization by an X-ray photon or a charged particle in a single thin layer would be subjected to large fluctuations \cite{pellecchia}; a laser test bench then allows the characterization of detectors with drift gaps thinner than \SI{500}{\micro\m}, such as an FTM instrumented with a high number of layers.

\subsection{Experimental setup}

The source in the optical setup is a pulsed UV laser of \SI{266}{\nano\m} wavelength, with average pulse duration of \SI{1}{\nano\s} and \SI{100}{\Hz} pulse repetition rate. The pulse energy is adjustable up to \SI{51}{\micro\J}, which allows a primary ionization cluster in the detector from a single up to 15 million photo-electrons. A single laser beam pulse releases electrons in the drift gap of the detector by two-photon ionization of low-potential impurity molecules naturally present in the gas in concentrations of some parts per million \cite{hilke_lasers}; for this reason, the rate of primary electrons as a function of the laser pulse energy is quadratic for low pulse energies, while it follows a linear trend at higher laser intensities.

\begin{figure}
    \centering
    \includegraphics[width=0.8\textwidth]{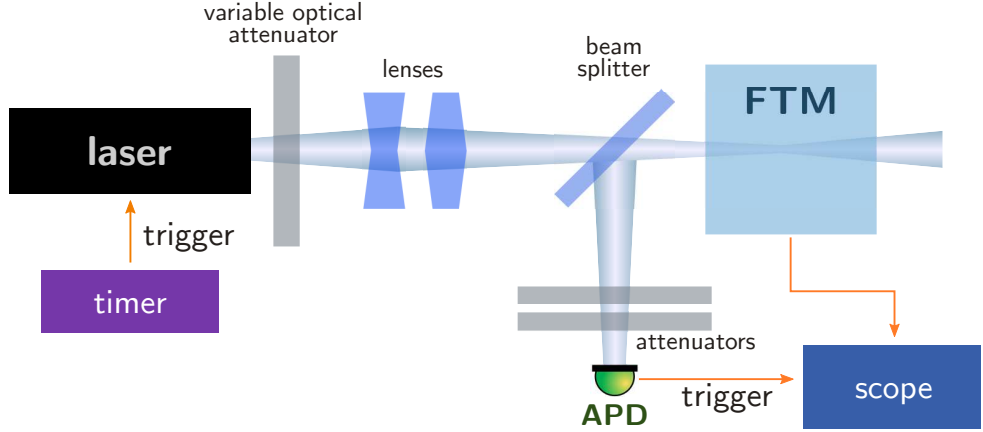}
    \caption{Sketch of the laser setup. The laser is triggered by a timer and the beam is attenuated by a first set of variable filters and focused; half of the beam is further attenuated and sent to an avalanche photodiode for triggering with time precision of the order of tens of picoseconds.}
    \label{fig:laser_setup}
\end{figure}

The full laser setup is depicted in Fig.~\ref{fig:laser_setup}. The laser beam is attenuated by a set of variable optical filters to adjust its intensity and focused by a set of lenses to a beam spot of \SI{25}{\micro\m}. The beam is divided by a 50\%-50\% beam splitter and is sent both to the detector and to an avalanche photodiode (Thorlabs APD430A) used as a trigger. The time resolution of the trigger APD was measured by irradiating in coincidence two identical APDs and found lower than \SI{50}{\pico\s}.

\subsection{Measurement technique}

The effective gain of the prototype has been measured in the laser setup instrumented with different amplification foils. The detector effective gain is defined as the ratio between the number of electrons collected from the \emph{ground} electrode and the number of primary ionization electrons and can be measured as $g_\text{eff} = i_\text{ground}/i_\text{primary}$.

The effective gain is measured as a function of the amplification voltage, i.e. the voltage difference between the anode and ground electrodes. At a fixed amplification voltage, the \emph{ground} current is measured with an ammeter of \SI{100}{fA} resolution. The primary current instead is measured in the primary ionization regime, i.e. by operating the detector at low amplification voltage such that the charge multiplication is equal to 1. In this field configuration, the primary ionization electrons are collected either from the \emph{anode} or from the \emph{ground} electrode; therefore, the primary current is measured as the sum of the anode and ground currents.

\subsection{Results}

\begin{figure}
    \begin{minipage}[c]{.58\textwidth}
        \includegraphics[width=1\textwidth]{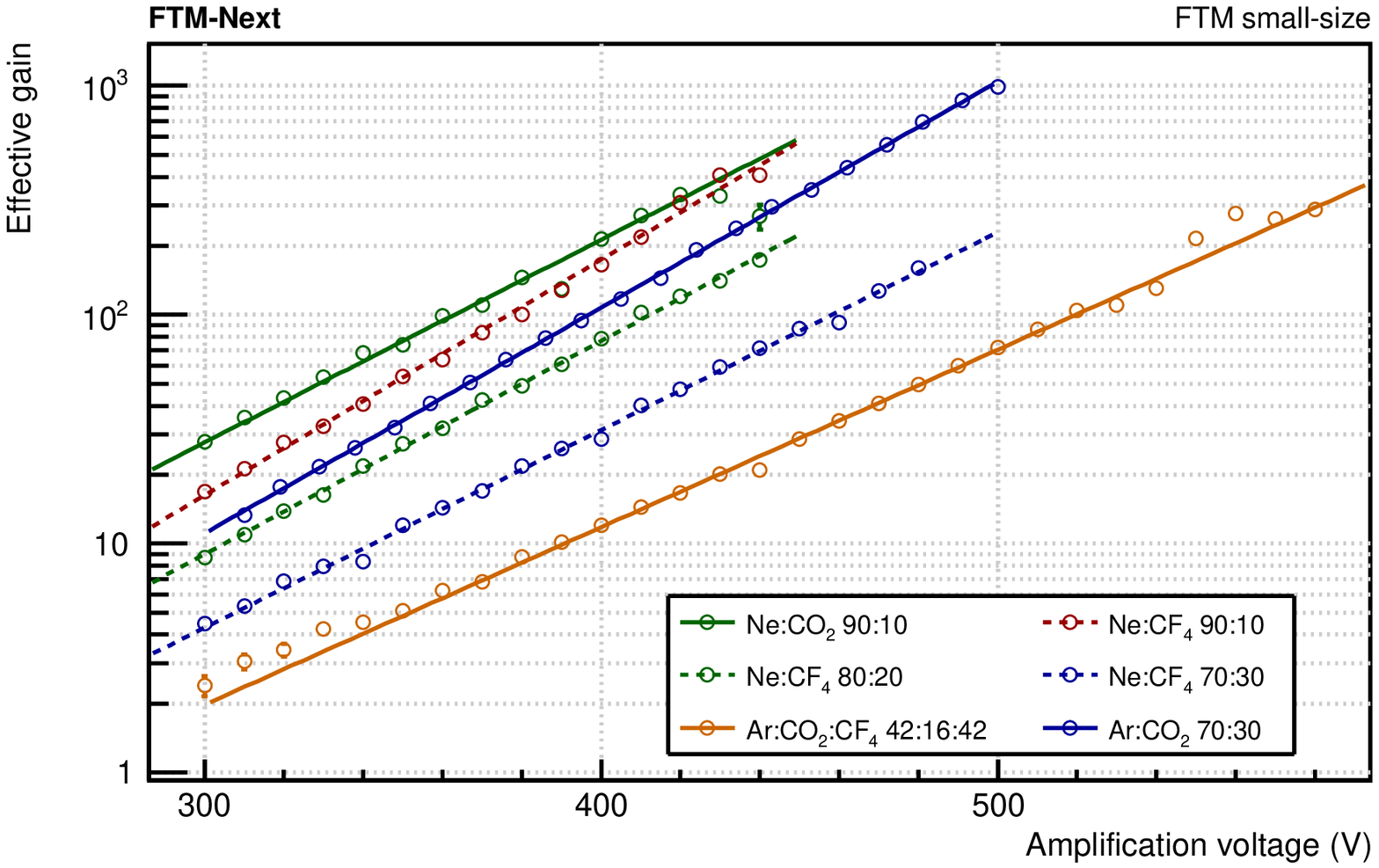}
        \caption{Comparison of effective gain curves obtained with \ce{CO2}- and \ce{CF4}-based mixtures.}
        \label{fig:gains}
    \end{minipage}
    \begin{minipage}[c]{.39\textwidth}
        \includegraphics[width=0.97\textwidth]{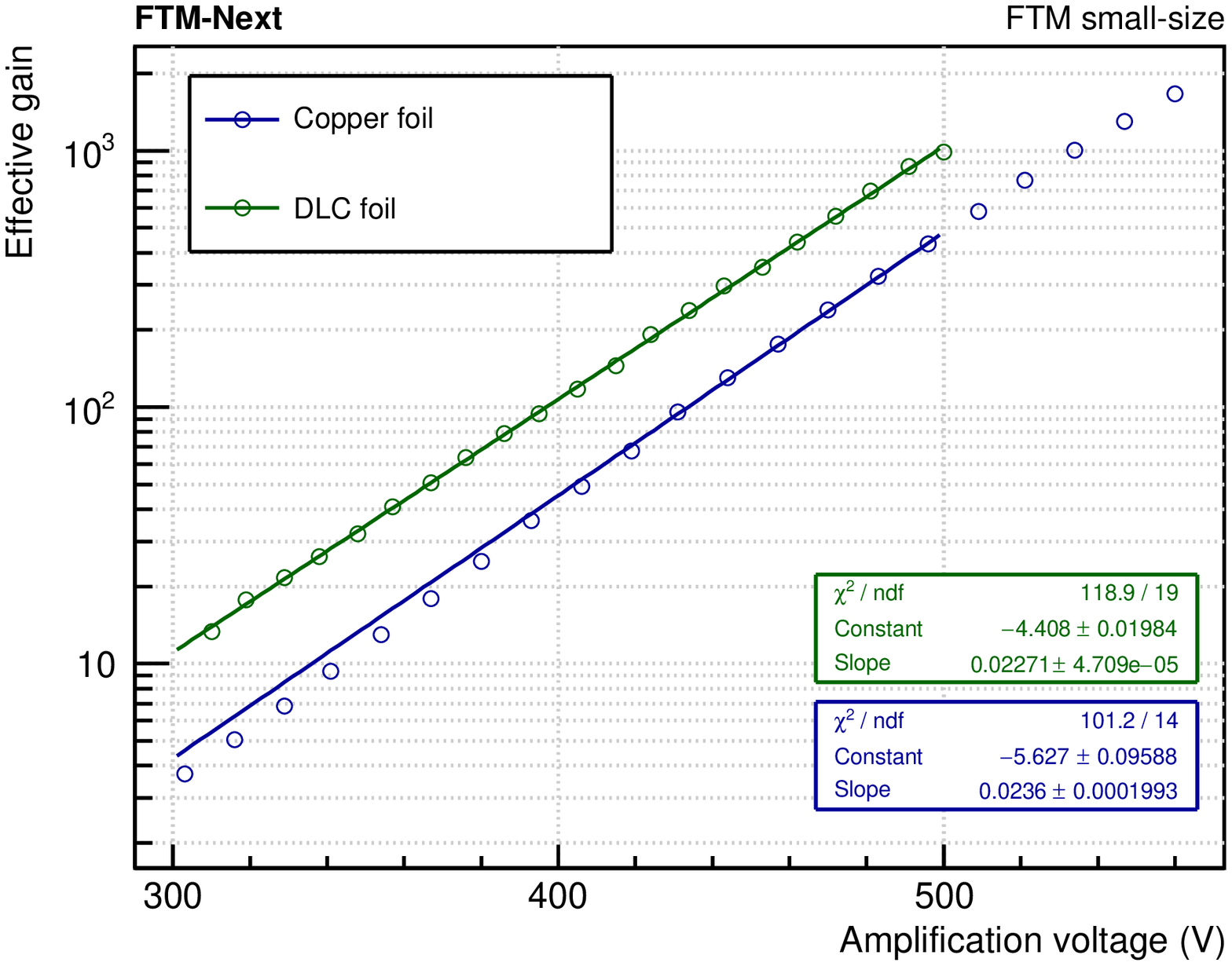}
        \caption{Effective gain of conductive and resistive amplification foils measured in Ar:\ce{CO2} 70\%:30\%.}
        \label{fig:gains_copper_dlc}
    \end{minipage}
\end{figure}

The gain measurement in the laser setup has been repeated on the prototype for several gas mixtures. Fig.~\ref{fig:gains} compares the gain curves as a function of the amplification voltages for Ar and Ne based mixtures. The highest gain obtainable with such mixtures before reaching the point of discharge of the detector is $10^3$ in Ar:\ce{CO2} 70\%:30\%, lower than the gain required for a detector fully efficient to MIPs.

To compare the results obtained with the FTM to a similar better-known detector, the µ-RWELL, the gain measurement was repeated after replacing the resistive foil with a copper-anode amplification foil. Fig.~\ref{fig:gains_copper_dlc} shows the results of a gain comparison between the resistive and conductive layouts in Ar:\ce{CO2} 70\%:30\%. At a fixed applied voltage, the resistive foil shows an effective gain equal to twice the gain of the conductive foil. A simulation study comparing the \emph{normal} and \emph{inverted} well geometries shows that this difference is exclusively to be ascribed to the difference in the hole shapes. However, the overall highest gain reachable with the conductive layout is higher than the gain in the resistive layout, because of the higher amplification voltage reachable before the onset of discharge. A similar behaviour is observed with other \ce{CO2} and \ce{CF4}-based gas mixtures.

\begin{figure}
    \centering
    \includegraphics[width=0.95\textwidth]{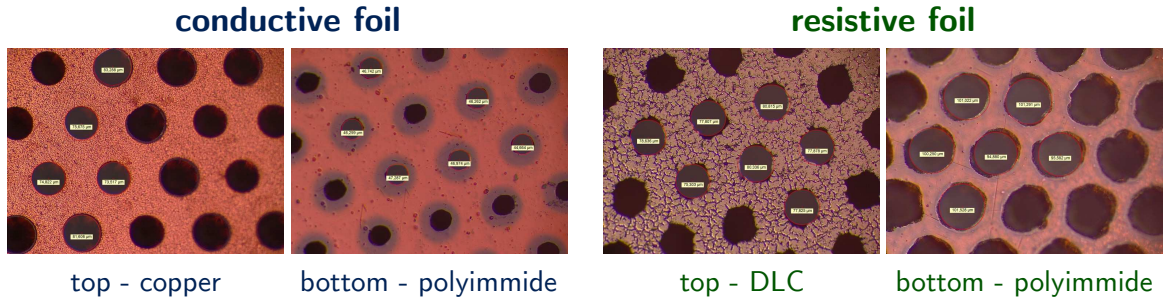}
    \caption{Conductive and resistive amplification foils observed at the microscope.}
    \label{fig:pics_microscope}
\end{figure}

An explanation for the lower stability to high fields for the resistive foils with respect to the conductive ones has been obtained by observing the hole pattern on the microscope (Fig.~\ref{fig:pics_microscope}). The two pictures on the left show the top and bottom sides (respectively made of copper and polyimide) of the conductive foils, with circular hole shapes of high regularity; by comparison, the resistive foil (in the two pictures on the right) shows an highly irregular hole shape not only on the top (DLC) side, but also on the bottom (polyimide) side. The resulting local irregularities in the electric field are associated with the increased discharge probability.

\begin{figure}
    \centering
    \includegraphics[width=0.95\textwidth]{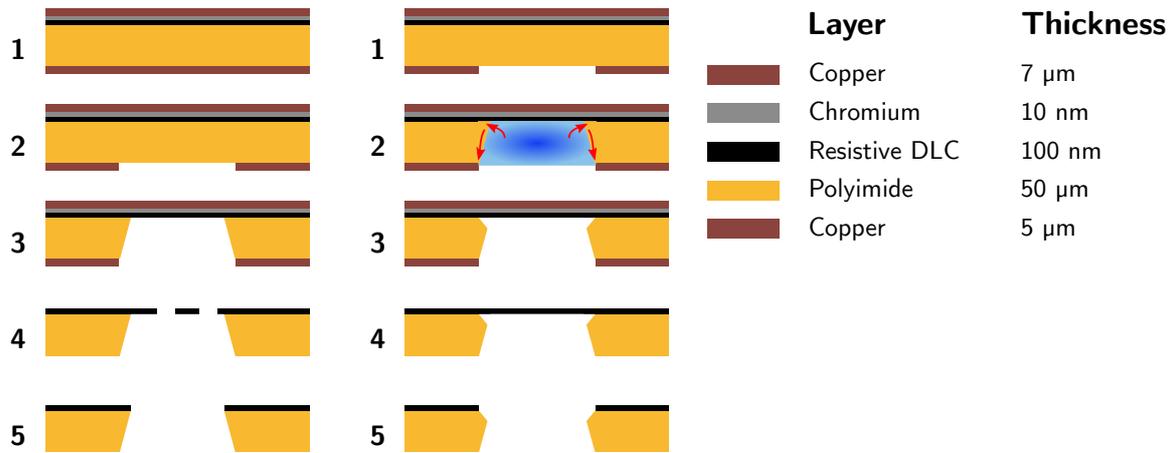}
    \caption{On the left, schematic steps of the production of resistive GEM foils from starting FCCL; on the right, production steps emphasizing the over-etching of the polyimmide resulting in hole walls irregularities.}
    \label{fig:steps_production}
\end{figure}

The origin of such DLC delamination has been pointed to one of the central steps of the foil production from the original laminate, as described in Fig.~\ref{fig:steps_production}; in particular, in the polyimide etching step the etching solution enters in contact with the DLC, infiltrating in the DLC itself due to imperfect adhesion in the polymide-DLC interface. This causes an over-etching of the polyimide holes resulting in irregular polyimide and DLC borders.

\begin{figure}
    \centering
    \includegraphics[width=0.9\textwidth]{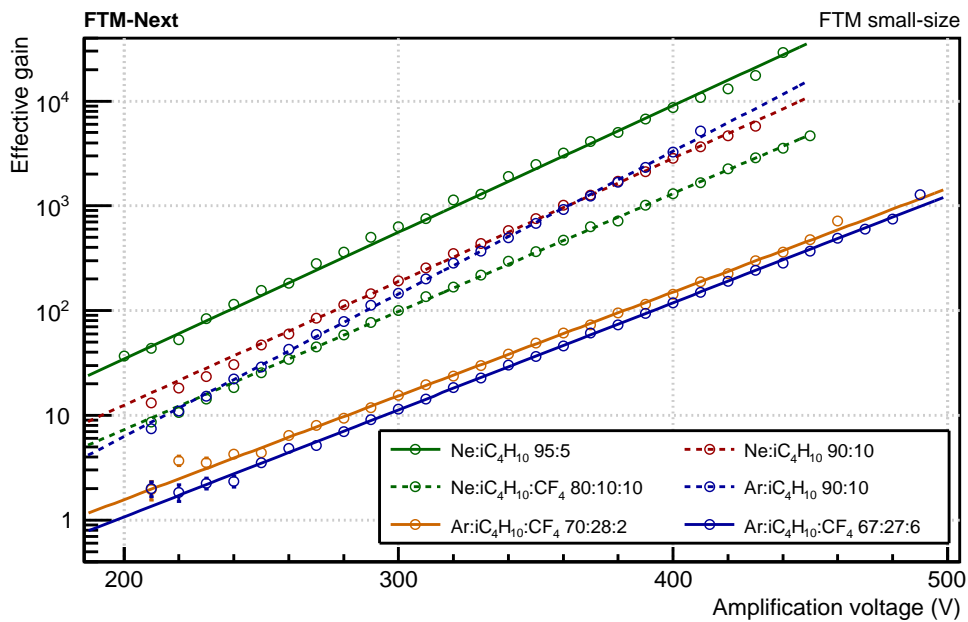}
    \caption{Comparison of effective gain curves obtained with isobuthane-based mixtures.}
    \label{fig:gains_isobuthane}
\end{figure}

A solution to the increased discharge probability is found by operating the detector with isobuthane-based mixtures. A comparison between such mixtures is shown in Fig.~\ref{fig:gains_isobuthane}, which proves the effective gain reachable in Ar and Ne mixed with \ce{iC4H10} is in general higher than the gain reachable with \ce{CO2} based mixtures. The highest gain measured is \SI{3e4}{} in Ne:\ce{iC4H10} 95\%:5\%. The same results cannot instead be obtained with \ce{CF4}-based mixtures, used as a secondary quencher both with isobuthane and with carbon dioxide (see Fig.~\ref{fig:gains}).

\section{Conclusions}

The construction, manufacturing and test of an FTM prototype has been discussed, together with a systematic foil performance comparison obtained with different gas mixtures. The use of an optical setup based on a UV laser allows fast and precise gain measurements, as well as efficiency measurements not obtainable with X-ray sources. A compared measurement shows that the overall highest gain reachable with DLC GEM foils manufactured from FCCL coated by present state-of-the-art magnetron sputtering is lower than the one obtainable with conductive foils used for the µ-RWELL detector. While isobuthane-based mixtures do allow reaching gains higher than \SI{1e4}{}, future steps in the development of the FTM will require an improvement in the foil production, starting from the achievement of better adhesion in the initial FCCL production. For instance, the DLC deposition by magnetron sputtering has been shown to yield samples with better adhesion if performed with an \emph{assisting} secondary ion beam \cite{piet_dlc}.

\smallskip


\section{References}

\medskip
\begin{thebibliography}{9}

\bibitem{rui_ftm}
    R. De Oliveira et al,
    \emph{A novel fast timing micropattern gaseous detector: FTM},
    arXiv:1503.05330

\bibitem{pellecchia}
    A. Pellecchia et al,
	\emph{A UV laser test bench for micro-pattern gaseous detectors},
	\emph{JINST} {\bf 15} (2020) C04011--C04011.

\bibitem{hilke_lasers}
    H. J. Hilke,
    \emph{Detector calibration with lasers - a review},
    \emph{Nucl. Instrum. Methods Phys. Res} {\bf A252} (1986) 169-179.

\bibitem{piet_dlc}
    P. Verwilligen et al,
    \emph{Diamond-like carbon for the fast timing MPGD},
    \emph{J. Phys.: Conf. Ser.} {\bf 1498} 012015 (2020).
    
\end{thebibliography}
\end{document}